\documentclass[12pt]{article}
\usepackage{amssymb,graphicx,amsmath,graphics,xfrac,url,subfigure}
\usepackage{authblk}
\begin{document}
	
\title{Scattering of a Klein-Gordon particle by a smooth barrier}
	
\author{Eduardo L\'opez}
\affil{School of Physical Sciences and Nanotechnology, Yachay Tech University, 100119 Urcuqu\'i, Ecuador}

\author{Clara Rojas}
\affil{School of Physical Sciences and Nanotechnology, Yachay Tech University, 100119 Urcuqu\'i, Ecuador\\
	\textit{crojas@yachaytech.edu.ec}}

\maketitle

\begin{abstract}
We present the study of the one-dimensional Klein-Gordon equation by a  smooth barrier. The scattering  solutions are given in terms of the  Whittaker $M_{\kappa,\mu}(x)$ function. The reflection and transmission coefficients are calculated in terms of the energy, the height and the smoothness of the potential barrier. For any value of the smoothness  parameter we observed transmission resonances.

\end{abstract}
	
\section{Introduction}

In this article we  computed the scattering solutions of the one-dimensional Klein-Gordon equation in  presence of a smooth barrier. This is a mathematical interesting problem because the solutions of the Klein-Gordon equation are given in terms of the  Whittaker $M_{\kappa,\mu}(x)$ function, whose asymptotic behavior is well-known.

This smooth barrier is a short-range potential which presents scattering states. This potential is interesting because varying the smoothness of the curve can be represented from the potential barrier to the cusp potential barrier, in all cases we observed transmission resonances. These potential barriers have applications in several topics of the solid state physics.

The Klein-Gordon equation is used to describe spin-0 particles. This equation have been widely study in the literature for different physical systems both time-independent \cite{rojas:2005,rojas:2006a, rojas:2006b,rojas:2007,villalba:2007,alpdogan:2013,rojas:2014a, rojas:2014b,hassanabadi:2014,ikot:2015,chabab:2016,di:2016,aquino:2019} and time-dependent \cite{rowan:1976, murai:2009,bohme:2013,kawamoto:2018} Klein-Gordon equation.
The analytical solution of the time-independent Klein-Gordon equation for different potentials has been caused of a lot of interest in recent years, for both bound states \cite{rojas:2006a, rojas:2006b,alpdogan:2013,hassanabadi:2014,di:2016} and scattering solutions \cite{rojas:2005,rojas:2007,villalba:2007,rojas:2014a, rojas:2014b,hassanabadi:2014,ikot:2015,chabab:2016,aquino:2019}. It has allowed the understanding of several physical phenomena of Relativistic Quantum Mechanics such as the antiparticle bound state \cite{schiff:1940,bawin:1974}, transmission resonances \cite{rojas:2005,rojas:2007,villalba:2007}, and superradiance  \cite{manogue:1988,rojas:2014a,molgado:2018,rojas:2019a}.

This article is organized as follow. In section 2, we present the one-dimensional Klein-Gordon equation. In section 3, we present the   smooth barrier. In section 4 we study the solutions for scattering states, and calculate the transmission and reflection coefficients. The discussion of our results are given in section 5. Finally, in section 6 we give the concluding remarks.

\section{The Klein-Gordon equation}

The Klein-Gordon equation for free particles, in natural units $\hbar=c=m=1$, is given by \cite{greiner:1987},

\begin{equation}
\label{KG}
\hat{p}^\mu\hat{p}_\mu \varphi = \varphi,
\end{equation}
being $\hat{p}^\mu=i\left(\dfrac{\partial}{\partial t},-\vec{\nabla}\right)$, Eq. \eqref{KG} becomes:

\begin{equation}
\left(\Box+1\right)\varphi=0.
\end{equation}

We need to solve the Klein-Gordon equation interacting with a spatially one-dimensional potential, then we start finding the form of the Klein-Gordon equation with the interaction of an electromagnetic field.

\bigskip
The electromagnetic field is described by the four-vector \cite{greiner:1987}:

\begin{eqnarray}
A^\mu&=&(A_0,\vec{A}), \\
A_\mu&=&\eta_{\mu\nu} A^\nu=(A_0,-\vec{A}),
\end{eqnarray}
where $\eta_{\mu\nu}=\textnormal{diag}(1,-1)$.

\bigskip
The minimal coupling of the electromagnetic field is expressed in the form,

\begin{eqnarray}
\hat{A}^\mu&\rightarrow&\hat{p}^\mu-e A^\mu,\\
\hat{A}_\mu&\rightarrow&\hat{p}_\mu-e A_\mu .
\end{eqnarray}

\bigskip
The one-dimensional Klein-Gordon equation  minimally coupled to a vector potential $A^{\mu}$
can be written as \cite{greiner:1987}:

\begin{equation}
\label{KG_gen}
(\hat{p}^\mu- e A^\mu)(\hat{p}_\mu- e A_\mu)\varphi=\varphi.
\end{equation}

\medskip
Consider a spatially one-dimensional potential $e A_0=V(x)$, $\vec{A}=0$, and a stationary solution of the Klein-Gordon equation  $\varphi(x,t)=\phi(x)e^{-i E t}$, Eq. \eqref{KG_gen} can be written as:

\begin{equation}
\label{KG_x}
\frac{\mathrm{d}^{2} \phi(x)}{\mathrm{d}x^{2}}+\left\{[E-V(x)]^{2}-1\right\}\phi(x)=0,
\end{equation}
where $E$ is the energy of the particle.

\section{The smooth barrier}

\begin{figure}[htbp]
	\begin{center}
	\includegraphics[scale=0.520]{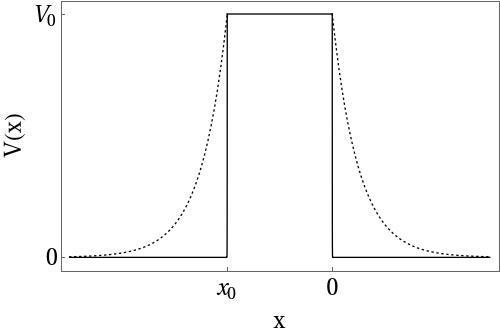}
	\end{center}
	\caption{\label{fig_pot}{Smooth barrier for $V_0=2$ with $a=0.5$ (dotted line) and square potential barrier for $a=0.001$ (solid line).}}
\end{figure}

The smooth barrier is given by:

\begin{equation}
\label{potential}
V(x) =
\left\{
\begin{array}{ll}
\quad	V_0 e^{\sfrac{(x-x_0)}{a}}, \quad \textnormal{for}  \quad x<x _0,\\
\quad V_0, \quad \textnormal{for} \quad x_0 \leq x\leq 0,\\
\quad	V_0 e^{-\sfrac{x}{a}}, \quad \textnormal{for} \quad  x > 0,
\end{array}
\right.
\end{equation}
where $V_0$ represents the height of the barrier and $a$ gives the smoothness of the curve. The form of the potential \eqref{potential} is showed in the Fig. (\ref{fig_pot}). From Fig. \ref{fig_pot} we can note that  the smooth barrier reduces to the square potential barrier for $a \rightarrow 0$. Also this potential reduces to the cusp potential barrier for $a \gg 0$ and $x_0=0$.

\bigskip

\section{Scattering States}

\subsection{Scattering solutions for $x<x_0$}

The scattering solutions for $x<x_0$ are obtained by solving the differential equation

\begin{equation}
\label{eq_x_I}
\frac{\mathrm{d}^2\phi_{\textnormal{I}}(x)}{\mathrm{d}x^2}+\left\{\left[E-V_0e^{\sfrac{(x-x_0)}{a}}\right]^2-1\right\}\phi_{\textnormal{I}}(x)=0.
\end{equation}

\medskip
On making the chang of variable $y=2iaV_0e^{\sfrac{(x-x_0)}{a}}$, Eq. (\ref{eq_x_I}) becomes

\begin{equation}
\label{eq_y_I}
y\frac{\mathrm{d}}{\mathrm{d}y}\left(y\frac{\phi_{\textnormal{I}}}{\mathrm{d}y}\right)-\left[\left(iaE-\sfrac{y}{2}\right)^2+a^2\right]\phi_{\textnormal{I}}=0.
\end{equation}

Putting $\phi_{\textnormal{I}}=y^{-\sfrac{1}{2}} f(y)$ we obtain the Whittaker differential equation

\begin{equation}
\label{f}
f(y)''+\left[-\frac 1 4 ´+\frac{iaE}{y} +\frac{\sfrac{1}{4}-a^2(1-E^2)}{y^2}\right]f(y)=0,
\end{equation}
which general solution is given by

\begin{equation}
\label{phi_y_I}
f(y)=c_1  M_{\kappa,\mu}(y)+d_1  W_{\kappa,\mu}(y),
\end{equation}
where $M_{\kappa,\mu}(y)$, $W_{\kappa,\mu}(y)$ are the Whittaker functions, $\kappa=iaE$ and $\mu=i\sqrt{E^2-1}a$. 
Then the solution to the Eq. \eqref{eq_y_I} is given by,

\begin{equation}
\label{phi_y_II}
\phi_{\textnormal{I}}(y)=c_1 y^{-\sfrac{1}{2}} M_{\kappa,\mu}(y)+d_1 y^{-\sfrac{1}{2}} W_{\kappa,\mu}(y).
\end{equation}

In terms of the variable $x$, Eq. \eqref{phi_y_II} becomes

\begin{eqnarray}
\label{phi_x_I}
\nonumber
\phi_{\textnormal{I}}(x)&=&c_1 (2ia V_0)^{(-\sfrac{1}{2})} e^{-\sfrac{(x-x_0)}{2a}} M_{\kappa,\mu}\left[2 i a V_0 e^{\sfrac{(x-x_0)}{a}}\right]\\
&+&d_1 (2ia V_0)^{(-\sfrac{1}{2})} e^{-\sfrac{(x-x_0)}{2a}} W_{\kappa,\mu}\left[2 i a V_0 e^{\sfrac{(x-x_0)}{a}}\right].
\end{eqnarray}

\bigskip
Because the asymptotic behavior of the Whittaker functions we only keep the solutions with the $M_{\kappa,\mu}(x)$  function. Then, from Eq. \eqref{phi_x_I}, the incident and reflected waves are,

\begin{eqnarray}
\label{phi_inc_ref}
\nonumber
\phi_{\textnormal{inc}}(x)&=&c_1 (2ia V_0)^{(-\sfrac{1}{2})} e^{-\sfrac{(x-x_0)}{2a}} M_{\kappa,\mu}\left[2 i a V_0 e^{\sfrac{(x-x_0)}{a}}\right],\\
\phi_{\textnormal{ref}}(x)&=&b_1 (2ia V_0)^{(-\sfrac{1}{2})} e^{-\sfrac{(x-x_0)}{2a}} M_{\kappa,-\mu}\left[2 i a V_0 e^{\sfrac{(x-x_0)}{a}}\right],
\end{eqnarray}
which are solutions of the differential equation \eqref{eq_x_I}.

\bigskip
\subsection{Scattering solutions for $x_0\leq x\leq 0$} 

The scattering solutions for $x_0<x<0$ are obtained by solving the differential equation

\begin{equation}
\label{x<x0and0}
\frac{\mathrm{d}^2\phi_{\textnormal{II}}(x)}{\mathrm{d}x^2}+\left[\left(E-V_0 \right)^2-1 \right]\phi_{\textnormal{II}}(x)=0,
\end{equation}

Eq. (\ref{x<x0and0}) has the general solution 

\begin{equation}
\label{phi_II}
\phi_{\textnormal{II}}(x)=b_2\,e^{-iqx}+c_2\,e^{iqx},
\end{equation}
where $q=\sqrt{\left(E-V_0\right)^2-1}$.

\bigskip
\subsection{Scattering solutions for $x >0$} 

The scattering solutions for $x>0$ are obtained by solving the differential equation

\begin{equation}
\label{eq_x_III}
\frac{\mathrm{d}^2\phi_{\textnormal{III}}(x)}{\mathrm{d}x^2}+\left[\left(E-V_0e^{-x/a}\right)^2-1\right]\phi_{\textnormal{III}}(x)=0.
\end{equation}

\medskip
On making the change of variable $z=2iaV_0e^{-\sfrac{x}{a}}$, Eq. (\ref{eq_x_III}) becomes

\begin{equation}
\label{eq_z_III}
z\frac{\mathrm{d}}{\mathrm{d}z}\left(z\frac{\phi_{\textnormal{III}}}{\mathrm{d}z}\right)-\left[\left(iaE-\sfrac{z}{2}\right)^2+a^2\right]\phi_{\textnormal{III}}=0.
\end{equation}

Putting $\phi_{\textnormal{III}}=z^{-\sfrac{1}{2}} g(z)$ we obtain the Whittaker differential equation

\begin{equation}
\label{g}
g(z)''+\left[-\frac 1 4 +\frac{iaE}{z} +\frac{\sfrac{1}{4}-a^2(1-E^2)}{z^2}\right]g(z)=0,
\end{equation}
which solution is given by

\begin{equation}
g(z)=c_3 M_{\kappa,-\mu}(z)+d_3W_{\kappa,-\mu}(z),
\end{equation}

\bigskip
Finally, the solution of Eq. \eqref{eq_z_III} becomes

\begin{equation}
\label{phi_y_3}
\phi_{\textnormal{III}}(z)=c_3 z^{-\sfrac{1}{2}} M_{\kappa,-\mu}(z)+d_3 z^{-\sfrac{1}{2}} W_{\kappa,-\mu}(z),
\end{equation}

\bigskip
In terms of the variable $x$, \eqref{phi_y_3} becomes

\begin{eqnarray}
\label{phi_x_III}
\nonumber
\phi_{\textnormal{III}}(x)&=&b_3 (2ia V_0)^{(-\sfrac{1}{2})} e^{\sfrac{x}{2a}} M_{\kappa,-\mu}\left(2 i a V_0 e^{\sfrac{-x}{a}}\right)\\
&+&d_3 (2ia V_0)^{(-\sfrac{1}{2})} e^{\sfrac{x}{2a}} W_{\kappa,-\mu}\left(2 i a V_0 e^{\sfrac{-x}{a}}\right).
\end{eqnarray}

From Eq. \eqref{phi_x_III} the transmitted wave is:

\begin{equation}
\label{phi_x_III}
\nonumber
\phi_{\textnormal{trans}}(x)=b_3 (2ia V_0)^{(-\sfrac{1}{2})} e^{\sfrac{x}{2a}} M_{\kappa,-\mu}\left(2 i a V_0 e^{\sfrac{-x}{a}}\right).
\end{equation}

\subsection{Transmission and reflection coefficients} 

For compute the transmission and reflection coefficient we need to use the definition of the electrical current.
The electrical current density for the one-dimensional Klein-Gordon equation (\ref{KG_x}) is defined as

\begin{equation}
\label{19}
j^\mu=\frac{i}{2}\left(\phi^*\,\partial^{\mu}\phi-\phi\,\partial^{\mu}\phi^*\right).
\end{equation}

The current as $x \rightarrow -\infty$ can be descomposed as $j_\textnormal{L}=j_\textnormal{in}-j_\textnormal{refl}$ where $j_\textnormal{in}$ is the incident current and $j_\textnormal{refl}$ is the reflected one. Analogounls we have that, on the right side, as $x \rightarrow \infty$ the current is $j_\textnormal{R}=j_\textnormal{trans}$, where $j_\textnormal{trans}$ is the transmitted current.

Using 
The reflection coefficient $R$, and the transmission coefficient $T$, are calculated by 

\begin{equation}
\label{R}
R=\frac{j_\textnormal{refl}}{j_\textnormal{inc}},
\end{equation}

\begin{equation}
\label{T}
T=\frac{j_\textnormal{trans}}{j_\textnormal{inc}},
\end{equation}
for which we need to identify the incident, reflected and transmitted wave. The quantities $R$ and $T$ are not independet, they are related via the unitary condition $R+T=1$.

Using the asymptotic behavior of the Whittaker function $M_{\kappa,\mu} \rightarrow e^{-\sfrac{y}{2}} y^{\sfrac{1}{2}+\mu}$, as $y\rightarrow 0$ \cite{abramowitz:1965}, we can write the incoming solution, the reflected solution and the transmitted solutions like a plane wave.
		
From Eq. \eqref{phi_x_I},  as $x \rightarrow -\infty$, the incident and reflected waves are given by

\begin{equation}
\label{phi_inc}
\phi_{\textnormal{inc}}(x)=c_1\,(2iaV_0)^{\mu}e^{i\sqrt{E^2-1}x},
\end{equation}

\begin{equation}
\label{phi_ref}
\phi_{\textnormal{ref}}(x)=b_1\,(2iaV_0)^{-\mu}e^{-i\sqrt{E^2-1}x},
\end{equation}
and from Eq. \eqref{phi_x_III} as $x \rightarrow \infty$ the transmitted  wave has the  form,

\begin{equation}
\label{phi_trans}
\phi_{\textnormal{trans}}(x) \rightarrow b_3\, (2iaV_0)^{-\mu} e^{i\sqrt{E^2-1} x}.
\end{equation}

Then 
\begin{equation}
\label{R_final}
R=\left| \dfrac{(2iaV_0)^{-\mu}}{(2iaV_0)^{\mu}}\right|^2\left|\dfrac{b_1}{c_1}\right|^2,
\end{equation}

\begin{equation}
\label{T_final}
T=\left| \dfrac{(2iaV_0)^{-\mu}}{(2iaV_0)^{\mu}}\right|^2\left|\dfrac{b_3}{c_1}\right|^2.
\end{equation}
In order to find $R$ and $T$, the  wave functions and their first derivatives must be matched at $x=x_0$ and $x=0$.   The coefficients $b_1$ and $b_3$ are calculated numerically in terms of $c_1$ from the following system of equations,

\begin{align}
&b_1\,M_{\kappa,\mu}(2iaV_0)+c_1\,M_{\kappa,-\mu}(2iaV_0)=b_2\, e^{-iqx_0}+c_2\, e^{iqx_0},\\
\nonumber
&b_1 \left[\left(-\dfrac{1}{2a}+i V_0+\dfrac{\kappa}{a}\right)  M_{\kappa,\mu}(2iaV_0)-\dfrac{1}{a} \left(\dfrac{1}{2}+\mu+\kappa\right)M_{\kappa+1,-\mu}(2iaV_0)\right]+\\
\nonumber
&c_1 \left[\left(-\dfrac{1}{2a}+i V_0-\dfrac{\kappa}{a}\right) M_{\kappa,-\mu}(2iaV_0)+\dfrac{1}{a} \left(\dfrac{1}{2}-\mu+\kappa\right)M_{\kappa+1,-\mu}(2iaV_0)\right]\\
&=- b_2\, q e^{-iqx_0}+c_2\, q e^{irx_0},\\
&b_2\,+c_2\, = b_3\, M_{\kappa,-\mu}(2iaV_0),\\
&i b_2\, q+i c_2\, q = b_3 \left[\left(\dfrac{1}{2a}-i V_0+\dfrac{\kappa}{a}\right)  M_{\kappa,-\mu}(2iaV_0)-\dfrac{1}{a} \left(\dfrac{1}{2}-\mu+\kappa\right)M_{\kappa+1,-\mu}(2iaV_0)\right].
\end{align}

\section{Results and discussion}

\begin{figure}[!th]
\centering
\includegraphics[scale=0.50]{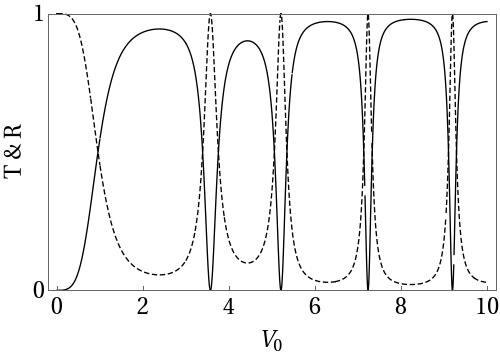}
\caption{Transmission and reflection coefficients in function of the height of  the barrier $V_0$ for $x_0=-1$, $a=0.5$ and $E=2$. The dashed line represents the transmission coefficient, and the solid line represents the reflection coefficient.}
\label{TR1}
\end{figure}

\begin{figure}[!th]
\centering
\includegraphics[scale=0.50]{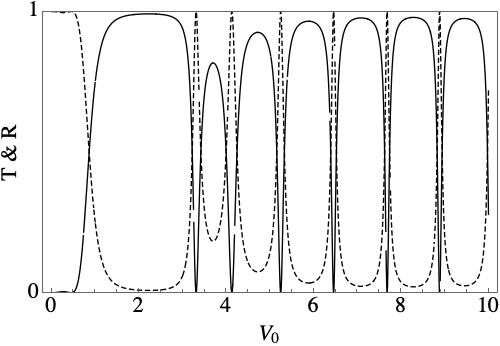}
\caption{Transmission and reflection coefficients in function of the height of the  barrier $V_0$ for $x_0=-2$, $a=0.5$ and $E=2$. The dashed line represents the transmission coefficient, and the solid line represents the reflection coefficient.}
\label{TR2}
\end{figure}

\begin{figure}[th!]
\centering
\includegraphics[scale=0.50]{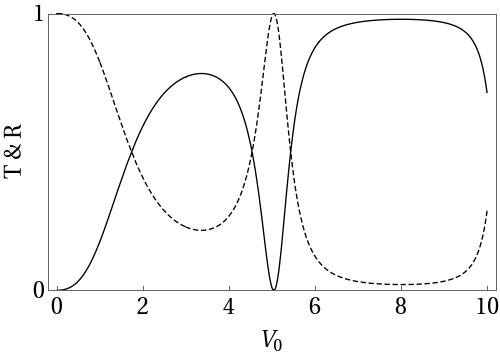}
\caption{Transmission and reflection coefficients in function of the height of the barrier $V_0$ for $x_0=0$, $a=0.5$ and $E=2$. The dashed line represents the transmission coefficient, and the solid line represents the reflection coefficient.}
\label{TR_cusp}
\end{figure}

\begin{figure}[th!]
\centering
\includegraphics[scale=0.50]{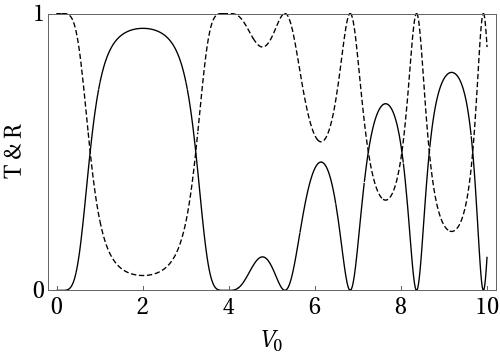}
\caption{Transmission and reflection coefficients in function of the height of the  barrier $V_0$ for $x_0=-2$, $a=0.001$ and $E=2$. The dashed line represents the transmission coefficient, and the solid line represents the reflection coefficient.}
\label{TR_barrier}
\end{figure}

Fig. (\ref{TR1}) shows the transmission and reflection coefficients for $x_0=-1$, $a=0.5$ and $E=2$ respect to the height of the barrier $V_0$. Fig. (\ref{TR2}) shows the transmission and reflection coefficients for $x_0=-2$, $a=0.5$ and $E=2$ respect to the height of the barrier $V_0$. When the value of $x_0$ increases, more peaks appear in the same range of $V_0$. Using the values $a=0.5$ with $x_0=0$ in Fig. (\ref{TR_cusp}) and $a=0.001$ in Fig. (\ref{TR_barrier})  we recover the results obtained by the cusp potential barrier and the square  potential barrier, respectively.  In all cases we observed transmission resonances. The relation $T+R=1$ also is satisfied. 

\bigskip

\bigskip

In the non-relativistic limit the Klein-Gordon equation reduces to the Sch\"odinger equation \cite{greiner:1987}. We wish to compare the scattering solutions of the Schrodinger equation with those of the Klein Gordon equation for the square potential barrier. 

For the square potential barrier showed in Fig. \ref{fig_pot} (solid line), the reflection $R$ and transmission $T$ coefficients are obtained by:

\begin{equation}
\label{Rbarrier}
R_\textnormal{barrier}=\dfrac{(k_1^2-k_2^2)^2 \sin^2\left(k_2 x_0\right)}{4 k_1^2 k_2^2+(k_1^2-k_2^2)^2\sin^2\left(k_2 x_0\right)}.
\end{equation}

\begin{equation}
\label{Tbarrier}
T_\textnormal{barrier}=\dfrac{4 k_1^2 k_2^2 }{4 k_1^2 k_2^2 +(k_1^2-k_2^2)^2\sin^2\left(k_2 x_0\right)},
\end{equation}
where  $k_1=\sqrt{2 E}$, $k_2=\sqrt{2 (E-V_0)}$ for the Sch\"odinger equation and  $k_1=\sqrt{E^2-1}$, $k_2=\sqrt{(E-V_0)^2-1}$ for the Klein-Gordon equation.

\begin{figure}[!th]
\begin{center}
\includegraphics[scale=0.45]{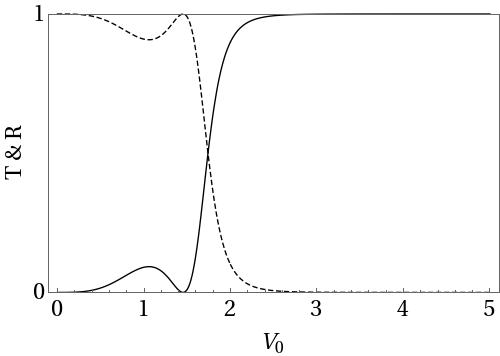}
\caption{\label{barrier_Schrodinger} The reflection $R$ and transmission $T$ coefficients varying energy $V_0$ of the  Sch\"odinger equation with the square potential barrier for $E=3$ and $x_0=-3$. The dashed line represents the transmission coefficient, and the solid line represents the reflection coefficient.}
\end{center}
\end{figure}

\begin{figure}[!th]
\begin{center}
\includegraphics[scale=0.45]{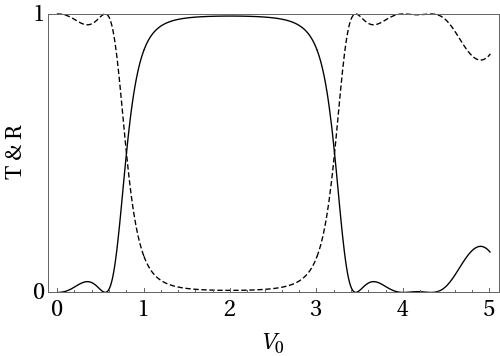}
\caption{\label{barrier_KG} The reflection $R$ and transmission $T$ coefficients varying energy $V_0$ of the Klein-Gordon equation with the square potential barrier for $E=3$ and $x_0=-3$. The dashed line represents the transmission coefficient, and the solid line represents the reflection coefficient.}
\end{center}
\end{figure}

In Figs. \ref{barrier_Schrodinger} and \ref{barrier_KG} we illustred the behaviour of the reflection and transmission coefficients in both cases. We can observed that for the Klein-Gordon equation more peaks appears in the transmission coefficient that for the Sch\"odinger equation. In both cases the relation $T+R=1$ is accomplished. 

\bigskip

\section{Conclusions}

In this paper we have studied the scattering solutions of the Klein-Gordon equation by a smooth barrier. We have calculated the transmission $T$ and reflection $R$ coefficients in function of the height of the potential $V_0$  for three different widths of the barrier. For certain values of the smoothness and weight of the potential barrier  we recover the results for the cusp potential \cite{rojas:2007} and for the square potential barrier \cite{greiner:1987,wachter:2010}.  In all cases we observe transmission resonances and the relationship $T+R=1$ is accomplished.  For future research we are going to consider the  bound-states solutions of this smooth potential barrier. 


\end{document}